# Overview of Screen Content Coding in Recently Developed Video Coding Standards

Xiaozhong Xu, *Member, IEEE*, Shan Liu, *Senior Member, IEEE*

*Abstract*—In recent years, screen content (SC) video including computer generated text, graphics and animations, have drawn more attention than ever, as many related applications become very popular. To address the need for efficient coding of such contents, a number of coding tools have been specifically developed and achieved great advances in terms of coding efficiency. The inclusion of screen content coding (SCC) features in all the recently developed video coding standards (namely, HEVC SCC, VVC, AVS3, AV1 and EVC) demonstrated the importance of supporting such features. This paper provides an overview and comparative study of screen content coding technologies, with discussions on the performance and complexity aspects for the tools developed in these standards.

## I. Introduction

IN the recently years, screen content video including computer generated text, graphics and animations, have drawn more attention than ever, as many related applications become very popular. However, conventional video codecs are typically designed to handle the camera-captured, natural video. Screen content video on the other hand, exhibits distinct signal characteristics and varied levels of the human's visual sensitivity to distortions. To address the need for efficient coding of such contents, a number of coding tools have been specifically developed and achieved great advances in terms of coding efficiency.

The importance of screen content applications is well addressed by the fact that all of the recently developed well-known video coding standards have included screen content coding (SCC) features. Nevertheless, the inclusion considerations of SCC tools in these standards [1] [2] [3] [4] [5] are quite different. Each standard typically adopts only a subset of the known tools in the category. Further, for one particular coding tool, when adopted in multiple standards, its technical features may various quite a lot from one standard to another.

All these caused confusions to both researchers who want to further explore SCC on top of the state-of-the-art and engineers who want to choose a codec particularly suitable for their targeted products. Information of SCC technologies in general and specific tool designs in these standards are of great interest.

This paper provides an overview and comparative study of screen content coding (SCC) technologies across a few recently developed video coding standards, namely HEVC, VVC, AVS3, AV1 and EVC. In addition to the technical introduction, discussions on the performance and design/implementation complication aspects of the SCC tools are followed up, aiming to provide a detailed and comprehensive report. The overall performances of these standards are also compared in the context of SCC.

The rest of this paper is organized as follows: Section II summarizes the existing screen content specific coding tools in various standards with comparative studies; Section III provides an overview of more generic coding technologies that has extra benefits on screen content materials; Section IV includes simulations results and discussion; The paper is concluded in Section V.

## II. Overview of Screen Content Specific Coding Tools

In this section, each of the following SCC specific tools will be introduced as a commonly acknowledged technology then followed by its standard specific technical variations. HEVC extensions on SCC is the first standard to include SCC features so features of coding tools included in this standard will be summarized as a basis to differentiate the same tool in other standards. Reasons behind the differences across different standards will also be elaborated.

### A. Intra Block Copy (IBC)

IBC is a block-based prediction technology, whose mechanism is similar to inter-picture motion compensation (MC). The essential difference lies in the fact that its reference samples are derived from inside the (reconstructed part of) current picture. It was also called current picture referencing (CPR) to reflect the similarity of IBC and inter motion compensation. The comparison of IBC and inter MC is shown

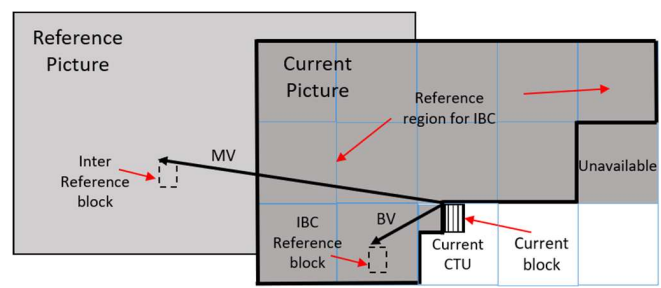

Fig. 1. Comparision of IBC and inter motion compensation

Xiaozhong Xu and Shan Liu are with Tencent America, 2747 Park Blvd, Palo Alto, CA 94306, USA (e-mail: {xiaozhongxu, shanl}@ tencent.com).

in Fig. 1. IBC was firstly proposed during the standardization of H.264/AVC [6]. It was formally included in HEVC SCC.

*Design of IBC Mode in HEVC SCC*

The IBC mode design in the HEVC SCC extensions [7] follows such a spirit that it is implemented almost in the same way as the HEVC inter-picture motion compensation, using the same syntax structure and nearly the same decoding process. In order to do that, the current (partially) decoded picture before the in-loop filtering process (including deblocking and SAO) is also regarded as a reference picture, when the IBC mode is enabled for coding of the current picture. In this way, block-based motion compensation and block-based intra sample copy are unified. For a coding block under inter mode, if its reference index points to the current decoded picture, the IBC mode is used [8]; otherwise, traditional inter mode is used.

The IBC reference picture in HEVC SCC is constrained, as shown in Fig. 1, where the white area is the not-yet-coded region of the current decoded picture; the samples in the grey area have already been reconstructed; the enclosed solid line inside this area establishes the available references for IBC mode, leaving out the top-right part of the reconstructed area relative to current CTU for parallel processing consideration. More details of IBC in HEVC SCC are described in [7].

*Implementation Implications*

On the implementation side, the addition of IBC mode has little impact to the existing software design, as there is no memory allocation distinction between storing the current decoding picture and other reference pictures. In addition, the decoding operation of IBC mode is very similar to the existing inter mode. However, for the hardware design, the above introduced full-frame based IBC mode does not merely require adding another reference picture.

In a typical hardware implementation of motion compensation (MC), the current decoded picture will be put to an off-chip memory location after all in-loop filtering operations, for future reference purpose. This operation is typically done piece by piece, such as in units of CTU. When an off-chip block in a reference picture is needed for MC for a current coding block, the related data including this reference block (and surrounding samples if interpolation is involved) will be fetched on-chip. This whole process involves memory bandwidth consumption in both writing and reading. Note that the memory bandwidth is considered to be very precious in a hardware system.

The behavior of the IBC mode in HEVC SCC is expected to be the same as the inter mode so that the existing inter MC module can be somehow reused. This is desirable as an extension to the main profile. However, there are two issues that differ the IBC mode from the inter MC:

- The current reference picture for IBC mode is an unfiltered version of the current decoding picture, which is different from the filtered picture that will be used for display or as a future reference picture. If the existing MC module is used to implement the IBC mode, this unfiltered picture needs to be put off-chip in addition to the filtered one. Writing this extra picture off-chip will cause the increase of memory bandwidth.

- A just-reconstructed neighboring block may probably become a reference for the next coding block coded in IBC mode. The timing for this case is too critical in that the whole process of "decoding a block - writing the reconstructed block off-chip - reading it back on-chip as a reference block" needs to be completed in a very limited number of cycles. This is considered impractical in typical hardware designs.

Due to the implementation difficulties, the scale of hardware-based deployment of HEVC SCC standard was limited. To facilitate with hardware implementations of IBC mode, adjustments have been made in later standards mainly to change the allowable IBC reference area in the current picture.

*Design of the IBC Mode in VVC*

*1) Reference range design*

In the implications mentioned earlier, allowing access to reference samples in a full-frame range will bring difficulties in cycling budget and increase in memory access bandwidth. The most noticeable difference applied in VVC for the IBC mode design is to constraint the allowed reference range to a local area. In this way, the needed reference samples will be stored in an on-chip memory. Both the memory bandwidth issue and implementation timing issue can therefore be resolved. In the followings, some detailed features of this mode in VVC are summarized.

In [9][10], the performance of allowing the current CTU as the search range was proposed, assuming only one CTU size of on-chip memory (referred as reference sample memory, or RSM) will be allocated for IBC prediction purpose. This constraint was the first adopted version of the IBC mode and considered as a compromise between coding efficiency and implementation cost. Later, a reuse mechanism of this RSM was introduced in [11], allowing some samples from coded CTU to the left to be used in the IBC mode.

The reuse method kept the required RSM size unchanged but can effectively improve the performance of the IBC mode. In Fig. 2, an example of such memory reuse mechanism is shown. At the beginning of each CTU, the RSM stores samples of the previous coded CTU (state (0)). When the current block is located in one of the four 64x64 regions in the current CTU, the corresponding region in the RSM will be emptied and used to store the samples of current 64x64 coding region. In this way, the samples in the RSM are gradually updated by the samples in the current CTU. Upon completion of the current CTU, the entire RSM is filled with all the samples of the current CTU (state (4)). In this example, the current CTU is partitioned firstly using quad-tree split. The coding order of the four 64x64 regions will then be top-left, top-right, bottom-left and bottom-right. In other block split decisions, the RSM update process is similar, to replace the respective regions in the RSM using the reconstructed samples in the current CTU. More details of RSM reuse design are provided in [12].

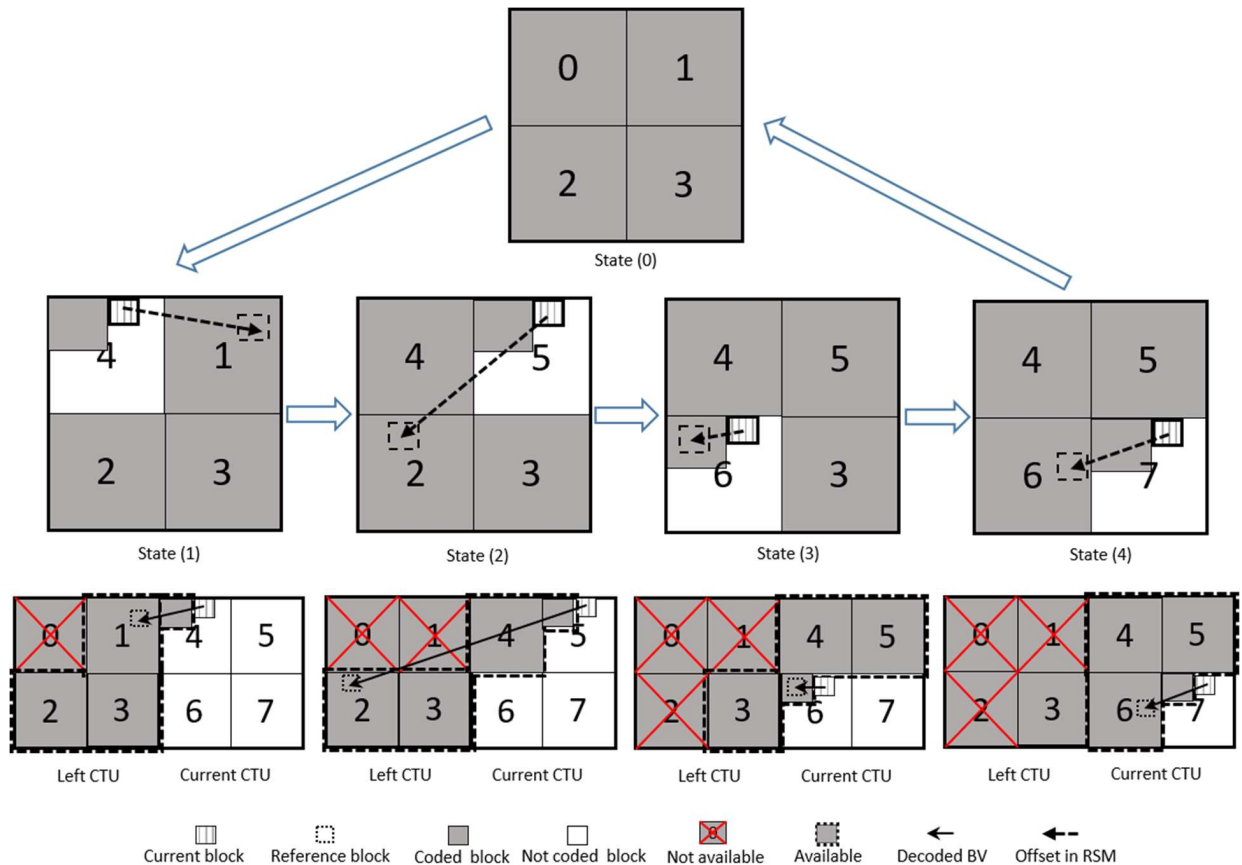

Fig. 2. Memory update process in the RSM during decoding of a CTU, quad-tree split is assumed at CTU root. Tow row: from RSM point of view; Bottom row: from picture point of view

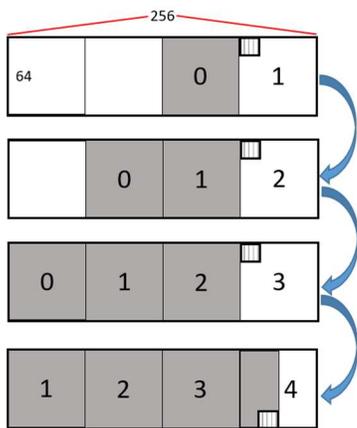

Fig. 3. Virtual buffer status update when CTU size is smaller than 128x128.

Note that to implement such a memory reuse mechanism as described above, the largest block size in IBC mode is limited as 64x64.

*2) Dual-tree structure handling*

In VVC, luma and chroma components can be coded separately, using different block partitioning structures. More specifically, a block partitioning structure shared by two chroma components is signaled independent of the one used for luma component in the CTU. This coding structure is referred as dual-tree structure. The separation is either from the root of a CTU (global dual-tree), which is enabled only for intra-coded slices [13], or from certain small block sizes (local dual-tree) when global dual-tree does not apply. The traditionally coding structure that includes a luma block and two corresponding chroma blocks is referred as joint-tree coding structure. Under the dual-tree structure, a chroma block's corresponding luma sample area may cover more than one luma block, therefore it can no longer derive its block vector from its collocated luma block. For simplicity, in such a case, only luma component can use IBC mode. Chroma components can be coded from one of available intra coding modes.

*3) Block vector coding*

IBC is no longer considered as part of inter mode in VVC standard. Rather, it is an independent coding mode, having its own vector coding engine as compared with the motion vector coding schemes in VVC inter mode [14][15]:

- A simplified merge candidate list design with 2 spatial neighboring blocks' BV and 5 history-based BV predictors (HBVP). Up to 6 candidates will be used in the list. The first two entries of the same predictor list will also be used for non-merge BV prediction mode.
- BV difference coding shares MV difference coding module.
- A module operation is applied to the decoded BV [16], after which the reference block will always be in either the current CTU or left CTU(s). In this way, arbitrary decoded BV values can be valid given the RSM reuse conditions are met.

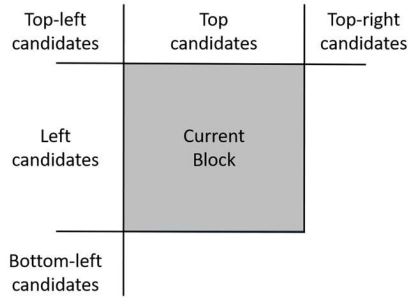

Fig. 4. Five spatial classes for IBC BV prediction in AVS3.

*Design of IBC Mode in EVC and AVS3*

The local-based IBC search range together with the RSM reuse mechanism as described in Fig. 3 has been adopted in EVC and AVS3 standards as well. Different from the BV coding in VVC, these two standards considered simplified BV coding mechanisms of their own.

In EVC, the BV coding reuses MVD coding module but without use of any BV prediction mechanism.

In AVS3, the concept of merge mode for IBC is removed. The BV predictor comes from "class-based" concept where a history-based BV list is first established and updated each time a block vector or string vector is coded [17]. In addition to record the vector value, the coded block's location, as well block size and occurrence frequency are also stored in the history table. For each IBC coded block, a CBVP table is built by classifying each entry in the history-based list into one of 7 classes, which consists of block size (greater than 32 samples), occurrence (more than 1 occurrence in history) and 5 locations (as shown in Fig. 4). An index is signalled to indicate which of the 7 classes is chosen. At the decoder side, the first entry in the chosen class will be used to predict the current block vector. The BVD coding part is the same as MVD coding.

*Design of IBC Mode in AV1*

The design of IBC mode in AV1 standard overcame the issues in HEVC SCC IBC implementation by imposing constraints in a way that the MC module can be reused for IBC mode without difficulty. With the below constraints, the procedure of IBC can be aligned with inter MC without apparent extra efforts.
- IBC is used only in intra coded pictures, and when IBC is turned on, the loop filters will be turned off. In this way, the memory bandwidth will not be increased.
- The allowed reference area for IBC includes the whole reconstructed part of current picture but excludes the current and left two CTUs. In this way, there will be enough time to write newly reconstructed reference samples off-chip, before it can be used for IBC reference.

B. *Palette Mode (PLT)*

The effectiveness of palette mode coding for screen content lies in the fact that in a local area, computer generated content may typically use a small number of colors to render the content. Therefore, instead of having the regular coding operations, coding these small color set directly would be more efficient. The colors to represent a coding block is therefore referred as color palette. Then each sample in the block is converted into an index of one entry in the palette. A typical PLT mode consists of representing the color palette and coding the index map. A PLT coded block does not have any residues. A color palette can be either joint palette or separate pallete. In the former case, a triplet—containing 1 luma value and two chroma values is used; For the later, the palette for luma is a single value and the one for two chroma components is a duplet.

*Design of PLT Mode in HEVC SCC*

In HEVC SCC [18], the entries of palette for the current block (up to 64) are joint triplets and come from two sources: reusing the palette predictor (up to 128) and decoding from the bitstream. A reuse flag for each predictor entry is signalled to indicate if this entry is used in the palette for current block. After that, a number of new entries will be signalled in the bistream. When completing the PLT coding, the palette predictor will be updated by 1) putting entries of current palette in front and 2) putting unused predictors from the previous palette predictor in the back until the maximum size has reached. An example of such predictor update process is shown in Fig. 6.

For index map coding, two primary modes are referred as copy-index (CI) and copy-above (CA). In the former mode, an index is signalled first, followed by a number of repeated index values along the scan direction; in the later mode, the current index is duplicated from the one at the same position of previous row (if horizontal traverse scan) or column (if vertical traverse scan). If a sample cannot be represented from the palette, it is called "escape" and its (quantized) value will be signalled directly. Note that "escape" pixel will be indicated using the largest index.

In terms of syntax arrangement, all indices at the start of an index run are coded together at the beginning, followed by a series of copy type/length combination. The values of all escape pixels are signalled at the end of this process. By doing the grouping, all the CABAC coded symbols (run type/length) are coded together so the parsing throughput can be improved. An example of a 4x4 PLT block coding with horizontal traverse scan order is shown in Fig. 7.

*Design of PLT Mode in VVC*

The palette mode in VVC is largely inherited from HEVC SCC. Some simplification modifications are made as follows:
- The entire block is divided into 16-sample coefficient groups (CG). Index map coding is processed CG by CG to reduce latency and memory requirement. A flag, run_copy_flag, is signaled for each sample in the CG to indicate if the copy mode is the same as the previous one.
- The maximum palette and palette predictor sizes are fixed to be 31/63, instead of signalling the sizes in SPS like in HEVC SCC.
- When CTU level dual-tree is used, luma and chroma coding are separate. In this case, the palette entry will contain only luma component or chroma components;
- When current block is under local dual tree, chroma CU does not use palette mode. For luma CU, it will update the palette predictor by putting some default values to the chroma components of new entries.
- The palette and predictor sizes are reduced by half when under either CTU level or local dual-tree.

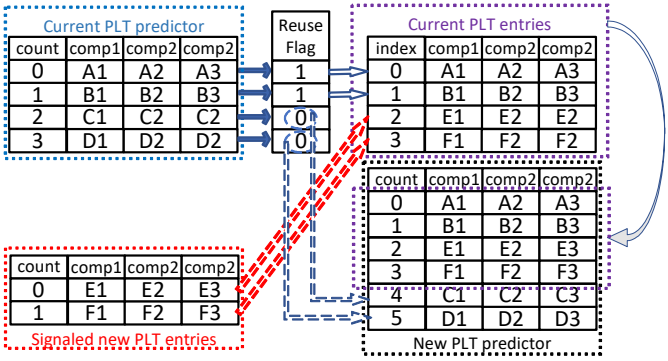

Fig. 5. Example of HEVC SCC PLT predictor update process.

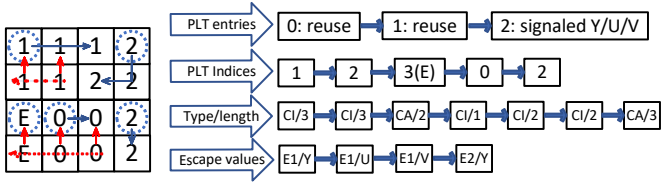

Fig. 6. Example of a 4x4 block under HEVC SCC PLT mode encoding flow.

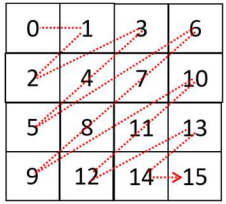

Fig. 7. Example of index scanning order in AV1 palette mode.

*Design of PLT Mode in AV1*

In AV1 [4], luma palette mode and chroma palette mode are determined independently so separate palettes are used. Up to 8 entries are allowed for each palette coded mode, which can be either predicted from neighbouring used palette colors or signalled for the delta part from the predictor.

The index map coding follows a diagonal scan order as shown in Fig. 8. For each index, it is coded using it top and left neighbouring indices (when available) as context.

Unlike in HEVC SCC and VVC, where no residue coding is applied to a palette coded block, transform coding and quantization is applied to the residue block in AV1 palette mode, just like the other intra prediction modes.

*C. Transform Skip Mode (TSM)*

Unlike the residue signal in camera captured contents, screen content residue signal tends to be in a low magnitude and scattered fashion. This behavior sometimes voids the assumption of the use of energy compacting transforms. Therefore, for screen content, the option of skipping transform coding may provide good coding performance improvement as compared to using transform for coefficient coding always.

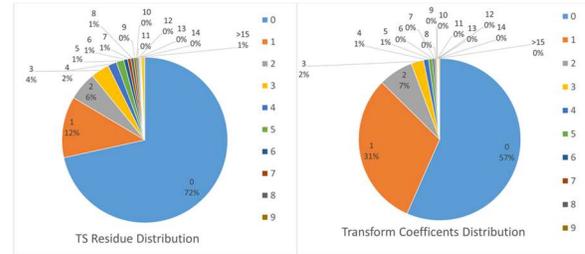

Fig. 8. Typical coefficient distributions of TSM and Transform blocks

*Design of TSM Mode in HEVC*

In HEVC version 1, TSM is enabled only for 4x4 blocks. In HEVC RExt [19] and SCC extensions, the allowed TSM sizes was extended to up to 32x32, which is also the maximum possible transform size. In all three standards, while the transform is skipped, the coding method of residue signals remains the same as that one designed for transform coefficients. In the recommended testing conditions of HEVC SCC, TSM was enabled only 4x4 blocks as increasing TSM block sizes did not prove to be a good performance/runtime trade-off using the reference software.

*Design of TSM Mode in VVC*

In VVC [20], a transform_skip_flag is signaled for coding block sizes up to 32x32. The residue coding of TSM is modified from regular transform coefficients. The motivation behind the specific design for TSM residue coding is from the observed distribution difference between a TSM block and a transform block, as shown in Fig. 8. By adjusting the coding engine, roughly additional 5% gain was reported. Some key features of the TSM residue coding in VVC are listed as follows:

- A TSM block will be divided into 4x4 sub-blocks.
- Forward scanning order applies among sub-blocks and coefficients within one sub-block.
- For each sub-block, coded_sub_block_flag is used to indicate if any non-zero residue exists.
- A sig_coeff_flag is used for each position along raster scan in the sub-block, to identify non-zero residues.
- Syntax flags abs_level_gtX_flag (X=1, 3, 5, 7, 9) are used to describe if the absolute value of each residue is greater than 1, 3, 5, 7, 9. par_level_flag is used to signal the parity of a residue when not greater than 9.
- abs_remainder flag to signal the leval if greater than 9.

*Design of TSM Mode in AVS3*

In AVS3 [21], a picture level flag is used to indicate if the TSM can be used. The block size applying TSM can go up to 32x32 as well. For the residue coding part, it is aligned with regular transform coefficient coding. The most apparent difference as compared to its counterpart in VVC, is the signalling of TSM usage. Instead of explicitly using a block level flag, the use of TSM in AVS3 is inferred by the parity of total number of decoded even coefficients in the block. In this way, if an encoder would like to choose TSM as compared to a regular transform (DCT-II), one of the coefficients may need to be adjusted by 1, if the existing number of even coefficients does not match the targeted parity for TSM. When TSM mode is not allowed for the picture, the same parity checking mechanism will be used to choose from different transform types.

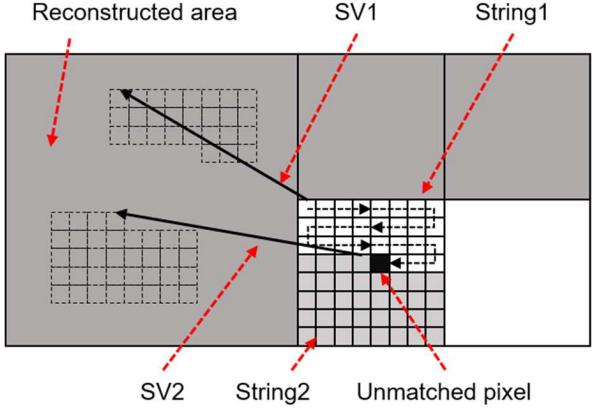

Fig. 9. Example of a block coded in ISC mode with two strings.

*D. Intra String Copy*

During the development of HEVC SCC Extensions, intra string copy (ISC) had been proposed and studied [22]. Similar to IBC, ISC is a predictive coding tool that explores the intra-picture similarity. A coding block is divided into several strings along the scanning direction. For each string, there are primarily two parameters to be signaled. The first one is a 2D string vector (SV) indicating the displacement between the current string and its reference. And the second one is the length of the string. An example of intra string copy mode is shown in Fig. 9. A simplified design of ISC mode has been later included in AVS3 standard. The detailed features of this mode in AVS3 [23] are summarized as follows:

- Reference sample memory and constraints

As previously discussed, the design of IBC mode has been improved after HEVC SCC by reducing the reference area from full-frame to a local area of the current and one left CTU. Further constraints are imposed such that all available reference samples can be stored in a 1-CTU size memory. In ISC mode, the reference memory access issue is similar as in IBC mode. In order not to increase the storage requirement, the available reference area for the ISC mode in AVS3 is unified with that of IBC mode. In other words, the same RSM is shared by both IBC and ISC modes. Further, some constraints are imposed to alleviate the implementation complexity:

1) Constraint on scanning direction: the memory block is typically allocated horizontally. In this context, ISC with a vertical scanning direction will have to make multiple memory accesses in order to acquire necessary reference pixels. To simplify the hardware implementation, currently only horizontal scanning direction is allowed in AVS3.

2) Constraint on strings: first, the reference string does not overlap with the current string. Further, the number of total strings allowed in a coding block is limited not to exceed N/4 (N is number of luma samples in the block). Third, the number of samples in each string should be a multiple of 4 so that the number of memory access can be reduced.

- SV prediction mechanism

Due to the similarity between IBC and ISC mode, the coding of the block vector (BV) and the string vector (SV) can be further improved by allowing cross-mode prediction between these two types of vectors. In IBC mode, A look-up table (History-based BV Predictor, or HBVP) is constructed to store the block vector of IBC mode as well as other related information (e.g. block size and location) in encoding or decoding order. This table is utilized to derive the predictor the block vector for each IBC coded block, using a class-based BV prediction (CBVP) method [17]. When ISC mode is enabled, the same HBVP table can serve the similar purpose for SV prediction as well by allowing the inclusion of coded string vectors into this table. The BV/SV prediction method involves the following technical characteristics:

1) Each time a coding block is encoded or decoded in ISC mode, the HBVP table is updated using the SVs of all the strings contained in the block; if a block is coded in IBC mode, the table is updated with current block's relevant information, as mentioned above.

2) In addition to recording the displacement vector information of the string, it is also necessary to record the position and size information of the string. Specifically, the position of the first pixel is taken as the position of the current string, and the size information is set to the length of the string.

3) For SV prediction in ISC mode, a flag is signaled first to indicate whether the SV of the current string can be found from the HBVP table. If yes, then the index of the selected entry of the table is signaled; otherwise, the SV is coded directly. For BV prediction in IBC mode, the process is unchanged.

*E. BDPCM*

The residues of an intra predicted block may still possess directional patterns since less correlation is maintained when the location of a to-be-predicted sample is moving away from the reference samples. To compensate such inefficiency, further prediction is applied among the residue samples. In BDPCM scheme of VVC [24], this mode is enabled at block level for intra coded blocks. Under BDPCM mode, a flag is used to choose the intra sample prediction as well as residue prediction from either horizontal or vertical direction. In equation (1) and (2), residue prediction of an MxN block along horizontal and vertical directions are shown, where $Q(r_{i,j})$ refers to quantized value of residue sample at position (i, j) of the block. The predicted residue is then to be further coded.

BDPCM can be turned on for luma and chroma components separately. When BDPCM flag for a block is equal to 1, TSM mode is inferred to apply without signaling.

$$\tilde{r}_{i,j} = \begin{cases} Q(r_{i,j}), & i = 0 \\ Q(r_{i,j}) - Q(r_{(i-1),j}), & 1 \leq i \leq (M-1) \end{cases} \quad (1)$$

$$\tilde{r}_{i,j} = \begin{cases} Q(r_{i,j}), & j = 0 \\ Q(r_{i,j}) - Q(r_{i,(j-1)}), & 1 \leq j \leq (N-1) \end{cases} \quad (2)$$

*F. Deblocking Modifications*

Intuitively, the mechanism of deblocking (smoothing one or more pixels near the block boundary) may not work well for screen contents, where sharp edges and abrupt changes can frequently occur within or across the block boundaries. It was reported that by disabling completely the deblocking operations for screen content video can produce some coding benefits. However, deblocking filters are typically control at picture level, making it difficult to handle videos with mixed screen and camera contents.

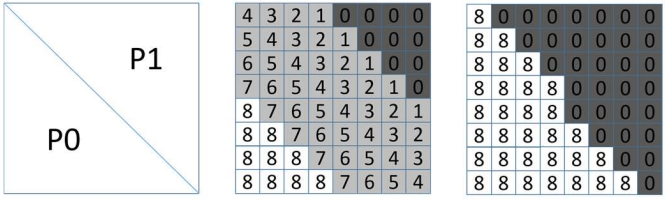

Fig. 10. Example of GPM mode (left), the blending on (middle) and off (right) at the partition boundary

Alternatively, the deblocking design in AVS3 [25] takes into consideration the properties of both types of contents by adding some screen-content-oriented pixel level measures. The key features of such additions in deciding the boundary strength (BS) are summarized as follows, reportedly the additional conditions do not negatively affect camera captured contents:
- Consider the cross-boundary difference. Typically for SCC, the sample difference between neighbouring pixels is either very large or very small. Before applying all the measures to decide the BS, one condition is checked that when the sample difference between two nearest samples at two sides of the edge is larger than a threshold, set BS=0; otherwise, other BS decision conditions apply.
- Consider the intra-region difference. During the BS decision process, a larger BS value reflecting stronger filtering needs, which will change more samples near the boundary while also taking more boundaries samples from the same side as input for filtering. If pixel difference among those boundary samples of the same side are larger than a threshold, a smaller BS value is chosen to lower the smoothing effect.

## III. Overview of Screen Content Related Technologies

In addition to the above introduced coding tools that are specially designed for screen contents, there are some generic coding tools but also show good benefits on screen contents. In this section, such coding tools in recently developed standards are discussed.

### A. Integer motion vector difference (IMVD)

Due to the nature of integer-based sample generation mechanism in most screen contents, the coded motion vector (MV) is likely to be in integer resolution as well. In this case, coding the MV difference with the default fractional-pel resolution becomes redundant. A slice level flag in HEVC SCC is used to indicate if all MVs inside the slice will be integer resolution [26]. When integer MV is enforced, the lower two bits of MV difference coding for fractional-pel part are skipped. However, the stored MVs (for future MV prediction and deblocking usage) are still in fractional-pel representation with their lower two bits being zero. The decoded integer MV difference should be left-shifted by two before adding to its predictor. It was reported that 3~4% gain can be achieved by this slice level adaptive resolution change for typical screen content video.

In VVC, the resolution decision has moved to block level when a decoded MV difference is not zero [27]. Further, when integer MV is used, it is followed by another flag to decide if 4 integer-pel is used to represent the MV resolution. This helps both camera-captured and screen contents by 1~2%.

Some similar approaches have also been implemented in EVC and AVS3, where an amvr_idex is signaled to choose among a set of potential vector resolutions (including integer-based ones). The same index also served as the decision in selection from a set of MV predictor candidates.

In VVC, a technique called MMVD (merge with MV difference) [28] is used by signaling a fixed set of delta values on top of existing inter merge candidates, as a short cut to code MV difference. The smallest delta value is ¼ sample in either x or y direction. Other delta values are designed as a multiple of it. To support efficient coding of screen content, where motion vectors are typically in integer precision, a slice level switch option is provided for MMVD. When enabled, all delta values are left shifted by 2 so no fractional delta values will be applied to the merge candidates. Roughly 1% coding gain for screen content was reported by using this shifting operation [29].

### B. Intra subblock partitioning (ISP)

In VVC, for intra coded luma blocks with sizes in the range from 8x4/4x8 to 64x64, the prediction block may be split equally into 4 (when larger than 8x4/4x8) or 2 prediction partitions along horizontal or vertical directions given that there is at least 4-sample width of a prediction sub-block [30]. As for transform sub-partitions, 1xN or 2xN blocks are allowed so there will always be four transform sub-blocks in this mode. For natural contents, the benefit of ISP mode is around 0.5% while for SCC contents, roughly 10% gain can be observed. The significantly higher gain probably comes from the nature of computer-generated patterns where neighboring samples can change dramatically so shorter distance of intra prediction reference can help to reduce the residue. In addition, smaller size transform can also help to concentrate the residue energies, which are not distributed evenly within a coding block.

### C. Geometrical partition mode blending off (GPMBO)

In order to capture motion object boundaries, block partitioning schemes are used to reduce to size of the coding block so that samples contained inside one block can share the same motion. Typically, the division is done by splitting a rectangular block into several smaller rectangular ones. However, motions do not always go with horizontal/vertical directions nor can be divided in the middle of a block. A more flexible partitioning scheme, geometrical partition mode (GPM), refers to applying a set of angles other than horizontal /vertical and/or unequal size partitioning in prediction block splitting. For camera-captured content, sample values near an object boundary usually transit smoothly from one to another. Therefore, a transition area across the partition boundary in GPM is designed, in which prediction samples are generated by using weighted average of the two partitions. In Fig. 10, an example of diagonal partition (into P0 and P1) with equal size is shown. For each partition P0 and P1, a reference block is assigned. It was reported that by applying the blending off mask, 1~2% coding gain can be achieved [31][32].

### D. Adaptive Color Transform (ACT)

The most common color format is YCbCr (or YUV) as 4:2:0 video is typically used in applications. In some high-end usage, especially related to screen content, RGB 4:4:4 video may be directly captured and coded. In-loop color transform in this case can show advantages and produce significant coding benefits.

In HEVC SCC, ACT [33] converts the original 4:4:4 color format into Y'CoCg format using the following formulas, (3) for lossy case and (4) for lossless case:

$$\begin{pmatrix} Y' \\ Co \\ Cg \end{pmatrix} = \begin{pmatrix} -1 & 2 & 1 \\ 2 & 0 & -2 \\ -1 & 2 & -1 \end{pmatrix} \begin{pmatrix} R' \\ G' \\ B' \end{pmatrix} \cdot \begin{pmatrix} 1/4 \\ 1/4 \\ 1/4 \end{pmatrix} \quad (3)$$

$$\begin{aligned} Co &= R' - B' \\ t &= B' + (C0 \gg 1) \\ Cg &= G' - t \\ Y &= t + (Cg \gg 1) \end{aligned} \quad (4)$$

In VVC, (4) is used in both cases [34]. It is reported that for RGB SCC contents, more than 8% coding gain can be achieved; while for YUV content, the benefit is limited, as expected.

### E. Hash based motion estimation (HashME)

In screen content video, object moving across pictures may not follow the optical flow model in camera-captured contents. As a result, the best matching reference for an inter coded block can be distant away from the collocated position in the reference picture, or even irrelevant to MV predictor indicated position. On the other hand, many repetitive textures exist from one picture to another. In light of these behaviors, a motion estimation method (referred as HashME) [35] has been developed by matching the current block's hash key with those in the reference pictures. This is far cheaper than conventional block-based matching ME since only positions with the same hash key need to be compared.

For each reference picture, a hash key is generated for each position. A hash table is then established as a linked list, containing positions of the same key. In hash based ME, the hash key of current block is calculated and only positions in the hash table of the same key will be compared. If a good match is found, conventional ME and other inter modes can be skipped. This encoder-only algorithm has been implemented in both HEVC SCC and VVC reference software and have achieved both significant coding performance improvements remarkable runtime reductions. For example, when HashME option is turned on, in RA configuration, ~5% BD rate reduction and ~15% runtime reduction can be achieved.

TABLE I   SCC FEATURES SUPPORTED IN STANDARDS

| Tools | HEVC SCC | VVC | AVS3 | AV1 | EVC |
|-------|----------|-----|------|-----|-----|
| IBC   | ✓        | ✓   | ✓    | ✓   | ✓   |
| PLT   | ✓        | ✓   |      | ✓   |     |
| TSM   | ✓        | ✓   | ✓    |     |     |
| BDPCM |          | ✓   |      |     |     |
| ISC   | studied  |     | ✓    |     |     |
| DBK   |          |     | ✓    |     |     |

TABLE II   TGM, GAMING AND NATURAL TEST SEQUENCES

| Sequences | Acronym | Resolution | Frame Rate | Bit Depth |
|-----------|---------|------------|------------|-----------|
| FlyingGraphics | FLY | 1920x1080 | 60 | 8 |
| Desktop | DKT | 1920x1080 | 60 | 8 |
| Console | CNS | 1920x1080 | 60 | 8 |
| ChineseDocument-Editing | CDE | 1920x1080 | 30 | 8 |
| EnglishDocument-Editing | EDE | 1920x1080 | 30 | 8 |
| Spreadsheet | SPS | 1920x1080 | 30 | 8 |
| BitstreamAnalyzer | BSA | 1920x1080 | 30 | 8 |
| CircuitLayout-Presentation | CLP | 1920x1080 | 30 | 8 |
| Program | PRG | 1920x1080 | 60 | 8 |
| WebEn | WBE | 1920x1080 | 60 | 8 |
| WordExcel | WDE | 1920x1080 | 60 | 8 |
| ArenaOfValor | AOV | 1920x1080 | 60 | 8 |
| BQTerrace | BQT | 1920x1080 | 60 | 8 |

## IV. SIMULATION RESULTS AND DISCUSSIONS

The adoption status of each discussed SCC specific coding tool is summarized in table I. In this section, a set of simulations are carried out to demonstrate the effectiveness of these SCC specific coding tools in the respective standards. To achieve this, anchors are selected as the encoders without using any SCC specific tools in section II for each standard. Results of combination of all coding tools are also shown, in addition to the effect of each coding tool. To further evaluate the overall SCC performance of each standard, all of them are compared to the HEVC encoder without SCC features (i.e. the HM software). For fair comparison, HashME in HEVC and VVC are always turned off as they are feasible but not implemented in other standards' reference software.

Encoders of each standard are selected as follows: HEVC reference software HM-16.20 [36]; HEVC SCC reference software SCM-8.6 [37]; VVC reference software VTM-10.0 [38]; AVS3 reference software HPM-9.0 [39]; AV1 reference software [40]; EVC reference software ETM-7.0 [41].

Across the board for testing conditions [42] [43] [44] [45] [46], AVS3 SCC common test condition (CTC) [46] seems to be the most versatile and content rich and is therefore selected for performing the tests in this paper. As shown in Table II, a total of 13 test sequences in 4:2:0 format are selected from the CTC. These sequences are divided into 3 categories, namely text and graphics with motion (TGM), gaming (G) and camera captured (CC). The TGM class represents the most typical type of screen content. Therefore, the coding performance on TGM class is the primary focus for evaluating the SCC coding tools, whereas coding performances on the other two classes are served as an auxiliary reference in the evaluation.

TABLE III SIMULATIONS RESULTS OF HEVC SCC FEATURES (SCC TOOLS OFF AS ANCHOR)

| Sequences | All Intra | | | | Random Access | | | |
|---|---|---|---|---|---|---|---|---|
| | IBC | PLT | IMVD* | Overall | IBC | PLT | IMVD* | Overall |
| **TGM AVE.** | 52.45% | 35.64% | - | 58.60% | 41.93% | 26.64% | 0.58% | 46.74% |
| AOV | 2.93% | 0.20% | - | 3.85% | 0.45% | -0.06% | 0.00% | 0.63% |
| BQT | 1.69% | -0.49% | - | 2.17% | -0.10% | -0.32% | 0.00% | 0.15% |

TABLE IV SIMULATIONS RESULTS OF VVC SCC FEATURES (SCC TOOLS OFF AS ANCHOR)

| Sequences | All Intra | | | | | Random Access | | | | |
|---|---|---|---|---|---|---|---|---|---|---|
| | IBC | PLT | TSM | BDPCM* | Overall | IBC | PLT | TSM | BDPCM* | Overall |
| **TGM AVE.** | 46.51% | 38.44% | 27.22% | 2.74% | 61.38% | 36.66% | 31.85% | 22.98% | 2.35% | 52.06% |
| AOV | 2.53% | 0.30% | 1.00% | 0.16% | 3.61% | 0.27% | 0.10% | 0.85% | 0.09% | 1.14% |
| BQT | 1.59% | -0.02% | 0.23% | 0.01% | 1.83% | 0.73% | 0.09% | 0.38% | -0.05% | 0.97% |

TABLE V SIMULATIONS RESULTS OF AVS3 SCC FEATURES (SCC TOOLS OFF AS ANCHOR)

| Sequences | All Intra | | | | | Random Access | | | | |
|---|---|---|---|---|---|---|---|---|---|---|
| | IBC | TSM | ISC* | DBK | Overall | IBC | TSM | ISC* | DBK | Overall |
| **TGM AVE.** | 46.43% | 17.08% | 19.32% | 2.42% | 64.37% | 35.19% | 13.68% | 15.37% | 5.75% | 55.16% |
| AOV | 1.71% | 0.00% | 0.00% | 0.00% | 1.71% | 0.21% | 0.00% | 0.00% | 0.00% | 0.19% |
| BQT | 0.86% | 0.00% | 0.00% | 0.00% | 0.86% | 0.34% | 0.00% | 0.00% | 0.00% | 0.33% |

TABLE VI SIMULATIONS RESULTS OF AV1 FEATURES (SCC TOOLS OFF AS ANCHOR)

| Sequences | All Intra | | | Random Access | | |
|---|---|---|---|---|---|---|
| | IBC | PLT | Overall | IBC | PLT | Overall |
| **TGM AVE.** | 43.92% | 26.54% | 54.01% | 26.98% | 17.81% | 35.30% |
| AOV | 0.00% | 0.00% | 0.00% | 0.00% | 0.00% | 0.00% |
| BQT | 0.00% | 0.00% | 0.00% | 0.00% | 0.00% | 0.00% |

TABLE VII SIMULATIONS RESULTS OF DIFFERENT STANDARDS WITHOUT SCC TOOLS, HM USED AS ANCHOR

| Sequences | SCM | | VTM | | HPM | | AV1 | | ETM | |
|---|---|---|---|---|---|---|---|---|---|---|
| | AI | RA | AI | RA | AI | RA | AI | RA | AI | RA |
| **TGM AVE.** | 1.66% | 1.66% | 11.30% | 16.64% | -13.25% | -0.22% | 17.12% | 45.60% | -22.00% | -9.08% |
| AOV | -0.55% | -0.17% | 25.30% | 34.12% | 16.41% | 25.23% | 12.37% | 26.43% | 10.31% | 23.44% |
| BQT | -0.48% | -0.36% | 17.96% | 32.03% | 13.41% | 25.74% | 10.63% | 42.21% | 6.44% | 18.73% |

TABLE VIII SIMULATIONS RESULTS OF DIFFERENT STANDARDS WITH SCC TOOLS, HM USED AS ANCHOR

| Sequences | SCM | | VTM | | HPM | | AV1 | | ETM | |
|---|---|---|---|---|---|---|---|---|---|---|
| | AI | RA | AI | RA | AI | RA | AI | RA | AI | RA |
| **TGM AVE.** | 58.60% | 46.74% | 65.84% | 60.72% | 59.90% | 55.87% | 61.99% | 63.62% | 30.37% | 23.09% |
| AOV | 3.85% | 0.63% | 28.04% | 34.80% | 17.87% | 25.39% | 12.37% | 26.43% | 12.67% | 23.48% |
| BQT | 2.17% | 0.15% | 19.50% | 32.70% | 14.21% | 25.98% | 10.63% | 42.21% | 8.55% | 19.19% |

The simulations involve the encoding and decoding of the test sequences under All Intra (AI) and Random Access (RA) configurations. BD-rate [47] difference is calculated from four QP points between the tested method and the anchor.

In table III, IV, V and VI, simulation results for individual coding tools inside HEVC SCC, VVC, AVS3 and AV1 are shown. In addition, cumulative gains by using all SCC tools of one standard are also demonstrated. The anchor in each table is generated by disabling all SCC tools in the respective standard reference software. For EVC codec, when IBC is enabled for ETM software, the coding gain for TGM class under AI/RA configurations are 42.7%/34.68%, respectively.

From these results, we can conclude that coding tools IBC, PLT, TSM, ISC can significantly improve the compression efficiency of SCC materials. In particular, IBC is the most effective one with more than 40% BD rate saving in AI configuration in every standard. The overall improvements of SCC features are beyond 60+% for VVC and AVS3. While for HEVC SCC and AV1 the numbers are also tremendous. Several notes here are: adaptive IMVD mode encoder design is based on hash information therefore can only be used when HashME is turned on; BDPCM mode is enabled only when TSM mode is enabled (syntax constraint); ISC mode encoder can be turned on only when IBC mode is enabled. Results of these modes (with * in the above tables) are generated by comparing to their respective dependent modes (using the results of dependent modes as anchor), rather than the SCC off anchor. Further, the individual gain of one tool may not be its best performance result. For example, when IBC is, the gain of using DBK modification is much higher than reported here.

In addition to evaluating individual SCC coding tools within a standard, it is also of interest to compare the performance of different standards. Generally speaking, using the standard's reference software to perform this task is difficult. During a standardization process of each standard, coding tools are developed mainly to optimize the selected common test conditions, which are different from one to another. The performance-complexity tradeoff selections on different software platform also vary quite significantly. The success of one video coding standard lies also in the real implementations, not only from its reference software. The comparisons in this paper may not be able to draw a concrete conclusion as to which of the standard is better than others.

Table VII and VIII demonstrate the compression capacity across different standard codecs. The anchor in these two tables are generated using HEVC v1's software (HM). In table VII, all SCC tools are turned off in the tested results, in order to evaluate the performance of all standards' common coding capability, or in other words the base platform without SCC. In this set of tests, AV1 platform (without SCC) seems to be the one with highest coding performance, especially for RA configuration. SCM without SCC is slightly better than HM mainly due to some encoder side optimizations. As for HPM and ETM, these two codes perform noticeably better for gaming and camera captured video but appear to be less efficient in SCC TGM class when compared with HM software.

In table VIII, all coding tools in respective standards are enabled, except for the HashME option in SCM and VTM. This tool is an encoder only method, which is feasible to all software but only implemented in SCM and VTM. Therefore, it is disabled to fairly compare the benefit brought up by normative coding. As a reference, additional 3% RA gain could be added to the numbers shown in the table. Compared to the famous reference (HEVC v1), all SCC enabled standards can significantly outperform the anchor without SCC. To be more specific, VVC and AV1 lead the race by a few percent above AVS3 and HEVC SCC. The numbers for EVC is less attractive due to limited SCC support (only IBC).

## V. CONCLUSIONS

In this paper, screen content coding technologies appeared in the recently developed video coding standards have been reviewed and discussed. With the presented powerful tools, the coding efficiency has been significantly improved on top of the existing benefits by generic new coding tools in the respective standards. In addition, by utilizing some of the generic coding tools properly, additional coding benefit can be achieved.